\documentclass[12pt]{article}
\usepackage{amssymb,slashed,cite,epsfig}
\newcommand{\ft}[2]{{\textstyle\frac{#1}{#2}}}

\def\trace{\mathop{\rm Tr}\nolimits}
\def\sign{\mathop{\rm sign}\nolimits}
\def\rme{{\rm e}}

\def\rmd{{\rm d}}
\newsavebox{\uuunit}
\sbox{\uuunit}
    {\setlength{\unitlength}{0.825em}
     \begin{picture}(0.6,0.7)
        \thinlines
        \put(0,0){\line(1,0){0.5}}
        \put(0.15,0){\line(0,1){0.7}}
        \put(0.35,0){\line(0,1){0.8}}
       \multiput(0.3,0.8)(-0.04,-0.02){12}{\rule{0.5pt}{0.5pt}}
     \end {picture}}
\newcommand {\unity}{\mathord{\!\usebox{\uuunit}}}
\newcommand{\QED}{{\hspace*{\fill}\rule{2mm}{2mm}}}

\csname @addtoreset\endcsname{equation}{section}

\newcommand{\SO}{\mathop{\rm SO}}
\newcommand{\U}{\mathop{\rm {}U}}

\def\CR{\mathbb{R}}
\def\CM{\mathcal{M}}
\def\CC{\mathcal{C}}

\def\avall{\vskip 0.3cm}
\def\espai{\;\;\;\;\;\;\;}
\def\zespai{\;\;\;\;}
\def\s{\sigma}
\def\8M{$\CM_8$}
\def\k{\kappa}
\def\be{\begin{equation}}
\def\ee{\end{equation}}
\def\G{\Gamma}
\def\g{\gamma}
\def\ei{e^{\underline{i}}}
\def\ej{e^{\underline{j}}}
\def\e1{e^{\underline{1}}}
\def\1u{\underline{1}}
\def\2u{\underline{2}}

\def\0u{\underline{0}}
\def\e{\epsilon}
\def\target{$\CR^{1,1}\times \mathcal{M}_8$ }
\def\target2{$\CR^{1,1}\times \mathcal{M}_8$,}
\def\9G{\G_{\underline{9}}}
\def\otaula{\begin{tabular}}
\def\ctaula{\end{tabular}}
\def\yvec{\vec{y}}

\def\undos{{1\over 2}}
\def\p{\partial}
\def\we{\wedge}
\def\te{\tilde{e}}
\def\t*{\tilde{*}}

\def\euc{\mathbb{E}}

\def\nn{\nonumber \\}
\def\bea{\begin{eqnarray}}
\def\eea{\end{eqnarray}}

\begin{document}

 \begin{titlepage}
\begin{flushright}
ECM-UB-03/12 \\
IFUM-756-FT\\
KUL-TF-2003/07\\
hep-th/0304210
\end{flushright}
\vspace{.5cm}
\begin{center}
\baselineskip=16pt {\LARGE    Supertubes in reduced holonomy manifolds
}\\
\vfill {\large Joaquim Gomis$^1$, Toni Mateos$^1$, Pedro J. Silva$^2$ and
Antoine Van Proeyen$^3$
} \\
\vfill
{\small $^1$ Departament ECM, Facultat de F{\'\i}sica, \\
Universitat de Barcelona, Institut de F{\'\i}sica d'Altes Energies and\\
CER for Astrophysics, Particle Physics and Cosmology, \\
Diagonal 047, E-08028 Barcelona, Spain
\\ \vspace{6pt}
$^2$  Dipartimento di Fisica, Universit{\`a} di Milano, via Celoria 16,
I-20133 Milano\\
and Istituto Nazionale di Fisica Nucleare (INFN) - Sezione di Milano,
Italy\\ \vspace{6pt}
 $^3$ Instituut voor Theoretische Fysica - Katholieke Universiteit Leuven
\\
Celestijnenlaan 200D B--3001 Leuven, Belgium
}
\end{center}
\vfill
\begin{center}
{\bf Abstract}
\end{center}
{\small We show that the supertube configurations exist in all
supersymmetric type IIA backgrounds which are purely geometrical and
which have, at least, one flat direction. In other words, they exist in
any spacetime of the form $\CR^{1,1} \times \CM_8$, with \8M any of the
usual reduced holonomy manifolds. These generalised supertubes preserve
1/4 of the supersymmetries preserved by the choice of the manifold \8M.
We also support this picture with the construction of their corresponding
family of IIA supergravity backgrounds preserving from 1/4 to 1/32 of the
total supercharges.} \vspace{2mm}
 \vfill \hrule width 3.cm
{\footnotesize \noindent E-mail: \texttt{gomis@ecm.ub.es,
tonim@ecm.ub.es,\\ pedro.silva@mi.infn.it,
Antoine.VanProeyen@fys.kuleuven.ac.be } }
\end{titlepage}
\tableofcontents{}
\newpage
\section{Introduction and Results}

The fact that D-branes couple to background fluxes can allow, under the
appropriate circumstances, a collection of D-branes to expand into
another brane of higher dimension. Also the inverse process is observed,
where higher dimensional D-branes collapse into smaller dimensional ones
or even into fundamental strings. Non-supersymmetric examples of such
configurations are the expansion of Born-Infeld
strings~\cite{Emparan:1998rt}, the dielectric branes~\cite{Myers:1999ps}
and the matrix string theory calculations
of~\cite{Schiappa:2000dv,Silva:2001ja}. More recent supersymmetric cases
have also been constructed, like the giant gravitons in AdS
spaces~\cite{McGreevy:2000cw,Grisaru:2000zn}. All these configurations
share the handicap that the perturbative quantisation of string theory is
still not possible due to the presence of  Ramond-Ramond fluxes.

Supertubes~\cite{Mateos:2001qs} are very different from the former cases
because they are expanded configurations that live in a completely flat
space, with all other background fields turned off. They correspond to a
bound state of D0-branes and fundamental strings that expand into a D2
with tubular shape due to the addition of angular momentum. Remarkably,
they also preserve 1/4 of the 32 supersymmetries of the Minkowski vacuum,
unlike some other similar (but non-supersymmetric) configurations that
were constructed in~\cite{Harmark:2000na}. Furthermore, the simplicity of
the background allowed for a perturbative string-theoretical study of the
supertube, beyond the probe or the supergravity
approximations~\cite{Mateos:2001pi}.

The purpose of this paper is to show that it is possible to generalise
the construction of the original supertube configurations to other purely
geometrical backgrounds, while still preserving some supersymmetry. This
generalisation consists on choosing a type IIA background of the form
$\CR^{1,1} \times \CM_8$, with $\CM_{8}$ a curved manifold. Since we do
not turn on any other supergravity field, supersymmetry restricts \8M to
be one of the usual manifolds with reduced holonomy~\cite{Berger}:

\begin{center}
\otaula {||c||c||} \hline \hline
 \8M  & Fraction of the 32 supersymmetries preserved     \\ \hline \hline
$\CR^4 \times CY_2$  &          1/2           \\ \hline
$CY_2 \times CY_2$  &          1/4           \\ \hline
$\CR^2 \times CY_3$  &          1/4           \\ \hline
$CY_4$  &        1/8           \\ \hline
$\CR \times G_2$  &     1/8           \\ \hline
Spin(7)  &      1/8           \\ \hline
Sp(2)  &        3/8           \\ \hline\hline
\ctaula
\end{center}

We will show that it is possible to supersymmetrically embed the
supertube in these backgrounds in such a way that its time and
longitudinal directions fill the $\CR^{1,1}$ factor, while its compact
direction can describe an arbitrary curve $\CC$ in \8M.

The problem will be analysed in two different descriptions. In the first
one, we will perform a worldvolume approach by considering a D2 probe in
these backgrounds with the mentioned embedding and with an
electromagnetic worldvolume gauge field corresponding to the threshold
bound state of D0/F1. With the knowledge of some general properties of
the Killing spinors of the \8M manifolds, it will be shown, using its
$\k$-symmetry, that the probe bosonic effective action is supersymmetric.
As in flat space supertubes, the only charges and projections involved
correspond to the D0-branes and the fundamental strings, while the D2
ones do not appear anywhere. This is why, in all cases, the preserved
amount of supersymmetry will be 1/4 of the fraction already preserved by
the choice of background.

Note that, in particular, this allows for configurations preserving a
single supercharge, as is shown in one of the examples of this work.%
\footnote{This is not in contradiction with the fact that the minimal
spinors in 2+1 dimensions have 2 independent components since, because of
the non-vanishing electromagnetic field, the theory on the worldvolume of
the D2 is not Lorentz invariant.} In the other example that we present,
we exploit the fact that the curve $\CC$ can now wind around the
non-trivial cycles that the \8M manifolds have, and construct a supertube
with cylindrical shape $\CR \times S^1$, with the $S^1$ wrapping one of
the non-trivial $S^2$ cycles of an ALE space. In the absence of D0 and F1
charges, $q_0$ and $q_s$ respectively, the $S^1$ is a collapsed point in
one of the poles of the $S^2$. As $|q_0 q_s|$ is increased, the $S^1$
slides down towards the equator. Unlike in flat space, here $|q_0 q_s|$
is bounded from above and it acquires its maximum value precisely when
the $S^1$ is a maximal circle inside the $S^2$.

The second approach will be a spacetime description, where the
back-reaction of the system will be taken into account, and we will be
able to describe the configuration by means of a supersymmetric solution
of type IIA supergravity, the low-energy effective theory of the closed
string sector. Such solutions can be obtained from the original ones,
found in~\cite{Emparan:2001ux}, by simply replacing the 8-dimensional
Euclidean space that appears in the metric by \8M. We will show that this
change is consistent with the supergravity equations of motion as long as
the various functions and one-forms that were harmonic in $\euc^8$ are
now harmonic in \8M. It will also be shown that the supergravity solution
preserves the same amount of supersymmetry that was found by the probe
analysis.

Physically, the construction of these generalised supertubes is possible
because the cancellation of the gravitational attraction by the angular
momentum is a {\it local} phenomenon. By choosing the worldvolume
electric field $E$ such that $E^2=1$, and an arbitrary non-zero magnetic
field $B$, the Poynting vector automatically acquires the required value
to prevent the collapse {\it at every point of $\CC$}. This remains true
even after the replacement of the space where $\CC$ lives from $\euc^8$
by a curved \8M.

This paper is organised as follows: in section~\ref{ss:WVanal} we analyse
the system where the D2-supertube probes the $\CR^{1,1} \times \CM_8$,
and prove that the effective worldvolume action for the D2 is
supersymmetric using the $\k$-symmetry. In section~\ref{ss:Hamilanal} we
perform the Hamiltonian analysis of the system. We show that the
supersymmetric embeddings minimise the energy for given D0 and F1
charges, showing that gravity is locally compensated by the Poynting
vector. In section~\ref{ss:examples} we give to examples in order to
clarify and illustrate these constructions. Section~\ref{ss:SGanal} is
devoted to the supergravity analysis of the generalised supertubes. We
prove there the supersymmetry from a spacetime point of view. Conclusions
are given in section~\ref{ss:concl}.

\section{Probe worldvolume analysis} \label{ss:WVanal}

In this section we will prove that the curved direction of a supertube
can live in any of the usual manifolds with reduced holonomy, while still
preserving some amount of supersymmetry. The analysis will be based on
the $\k$-symmetry properties of the bosonic worldvolume action, and its
relation with the supersymmetry transformation of the background fields.

\subsection{The setup}

As announced, we consider a general IIA background of the form $\CR^{1,1}
\times \CM_8$, with $\CM_{8}$ a possibly curved manifold. In the absence
of fluxes, the requirement that the background preserves some
supersymmetry\footnote{In~\cite{Park:2003zn}, a first attempt to
construct supertubes in curved spaces was performed. Their configurations
are not supersymmetric because the backgrounds already destroy all
supersymmetries.} implies that $\CM_8$ must admit covariantly constant
spinors and, therefore, a holonomy group smaller than $\SO(8)$. The
classification of such manifolds is well-known~\cite{Berger}, and the
only possible choices for \8M are shown in the table of the introduction.

Let us write the target space metric as
\begin{equation}
\rmd s^2_{IIA}=-(\rmd x^0)^2+(\rmd x^1)^2+ \ei \ej
\delta_{\underline{ij}}\,, \espai \ei=\rmd y^j e_j{}^{\underline{i}} \,,
\espai i,j = 2,3,...,9\,,
\end{equation}
where $\ei$ is the vielbein of a Ricci-flat metric on \8M.
Underlined indices refer to tangent space objects.
We will embed the supertube in such a way that its time and
longitudinal directions live in $\CR^{1,1}$ while its
curved direction describes an arbitrary curve $\CC$ in $\CM_8$.
By naming the D2 worldvolume coordinates $\{\s^0,\s^1,\s^2\}$,
such an embedding is determined by
\begin{equation} \label{embed}
x^0=\s^0, \espai\espai x^1=\s^1, \espai \espai y^i=y^i(\s^2)\,,
\end{equation}
where $y^i$ are arbitrary functions of $\s^2$. The assignment of $\sigma
^0$ and $\sigma ^1$, i.e. the fact that $y^i$ is independent of $\sigma
^0$ and $\sigma ^1$, is a choice of parametrization.\footnote{In this
sense, the apparently rotating supertubes considered in~\cite{Cho:2003jd}
are indeed equivalent, through a worldvolume reparametrisation, to the
ordinary supertubes in flat space.}

Let us remark that, in general, the curve $\CC$ will be contractible in
\8M. As a consequence, due to gravitational self-attraction, the compact
direction of the D2 will naturally tend to collapse to a point.

Following~\cite{Mateos:2001qs}, we will stabilise the D2 by turning on an
electromagnetic flux in its worldvolume
\begin{equation} \label{fluxes}
 F_{\it 2}=E\, \rmd \sigma^0 \wedge \rmd \sigma^1
 + B\, \rmd \sigma^1 \wedge \rmd \sigma^2\,,
\end{equation}
which will provide the necessary centrifugal force to compensate the
gravitational attraction. In this paper we will restrict to static
configurations.

The effective action of the D2 is the DBI action (the Wess-Zumino term
vanishes in our purely geometrical backgrounds),
\begin{equation} \label{delta}
S= \int_{\CR^{1,1} \times C} \rmd \sigma^0 \rmd \sigma^1 \rmd \sigma^2
{\cal L}_{DBI}\,, \espai \espai {\cal L}_{DBI}=-\Delta\equiv
-\sqrt{-\det[g+F]}\,,
\end{equation}
where $g$ is the induced metric determined by the embedding $x^M(\sigma
^\mu )$,
and $F_{\mu \nu }$ is the electromagnetic field strength. $M$ denotes the
spacetime components $0,1,\ldots ,9$, and $\mu $ labels the worldvolume
coordinates $\mu =0,1,2$. The $\k$-symmetry imposes restrictions on the
background supersymmetry transformation when only worldvolume bosonic
configurations are considered. Basically we get $\Gamma _\kappa \epsilon
=\epsilon$ (see e.g.~\cite{Bergshoeff:1997kr}), where $\epsilon$ is the
background Killing spinor and $\Gamma_\kappa$ (see
e.g.~\cite{Bergshoeff:1997tu}) is a matrix that squares to~1:
\begin{equation}
\rmd^3\sigma \;  \Gamma _{\kappa }=\Delta ^{-1}\left[ \gamma _{\it
3}+\gamma _{\it 1} \Gamma
  _*\wedge F_{\it 2}\right].
 \label{Gammakappa}
\end{equation}
Here $\G_*$ is the chirality matrix in ten dimensions (in our conventions
it squares to one), and the other definitions are
\begin{eqnarray}
 \gamma _{\it 3} & = & \rmd\sigma ^0\wedge \rmd\sigma ^1\wedge \rmd\sigma ^2
 \,\partial _0x^M \partial _1x^N\partial _2x^P
 e_M{}^{\underline{M}}e_N{}^{\underline{N}}
 e_P{}^{\underline{P}}\Gamma_{\underline{MNP}}\,,
\nonumber\\
 \gamma _{\it 1} & = & \rmd \sigma ^\mu \partial _\mu
 x^Me_M{}^{\underline{M}}\Gamma_{\underline{M}}  \,.
 \label{hulp}
\end{eqnarray}
where $e_M{}^{\underline{M}}$ are the vielbeins of the target space and
$\Gamma_{\underline{M}}$ are the flat gamma matrices. We are using Greek
letters for worldvolume indices and Latin characters for the target
space.

We are now ready to see under which circumstances can the
configuration~(\ref{embed}),~(\ref{fluxes}) be supersymmetric. This is
determined by the condition for $\k$-symmetry, which becomes
\begin{equation} \label{ksym}
[\G_{\0u\1u}\g_{2} + E \g_2 \G_* +B \G_{\underline{0}}\Gamma
_*-\Delta]\e=0\,,
\end{equation}
where
\begin{equation}
\Delta^2=B^2+y'^{\underline{i}}y'^{\underline{i}}(1-E^2)\,, \qquad
y'^{\underline{i}}=y'^i e_i{}^{\underline{i}}\,,\qquad \gamma
_2=y'^{\underline{i}}\Gamma _{\underline{i}}\,, \espai\espai  y'^i
:=\partial_2y^i\,. \label{gamma2}
\end{equation}
The solutions of~(\ref{ksym}) for $\e$ are the Killing spinors of the
background, determining the remaining supersymmetry.

\subsection{Proof of worldvolume supersymmetry}
\label{ss:proof}

In this section we shall prove that the previous configurations always
preserve $1/4$ of the remaining background supersymmetries preserved by
the choice of \8M. We will show that the usual supertube projections are
necessary and sufficient in all cases except when we do not require that
the curve $\CC$ is arbitrary and it lies completely within the flat
directions that \8M may have. Therefore we first discuss the arbitrary
case, and after that, we deal with the special situation.

\avall

{\bf Arbitrary Curve:} If we demand that the configuration is
supersymmetric for any arbitrary curve in \8M, then all the terms
in~(\ref{ksym}) that contain the derivatives $y'^i(\s^2)$ must vanish
independently of those that do not contain them. The vanishing of the
first ones (those containing $\gamma _2$) give
\begin{equation} \label{F1}
\G_{\0u \1u }\Gamma _*\e=-E \e \espai \Longrightarrow \espai E^2=1\,,
\espai \textrm{and} \espai \G_{\0u \1u}\G_{*}\e=-\sign(E) \epsilon\,,
\end{equation}
which signals the presence of fundamental strings in the
longitudinal direction of the tube. Now, when $E^2=1$,
then $\Delta=|B|$, and the vanishing of the terms independent
of $y'^i(\s^2)$ in~(\ref{ksym}) give
\begin{equation} \label{D0}
\G_{\0u}\Gamma _*\e=\sign(B) \epsilon\,,
\end{equation}
which signals the presence of D0 branes dissolved in the worldvolume of
the supertube. Since both projections, (\ref{F1}) and (\ref{D0}),
commute, the configuration will preserve $1/4$ of the background
supersymmetries {\it as  long as they also commute with all the
projections imposed by the background itself.}

It is easy to prove that this will always be the case. Since the
target space is of the form \target2 the only nontrivial conditions that
its Killing spinors have to fulfil are
\begin{equation} \label{constant}
\nabla_i \e=\left(\partial_i + {1\over 4} w_i{}^{\underline{jk}}
\G_{\underline{jk}} \right)\e=0\,,
\end{equation}
with all indices only on \8M (which in our ordering, means $2 \leq i
\leq 9$). If one prefers, the integrability
condition can be written as
\begin{equation} \label{integrability}
[\nabla_i,\nabla_j]\e = {1 \over 4} R_{ij}{}^{\underline{kl}}
\G_{\underline{kl}}\e =0\,.
\end{equation}
In either form, all the conditions on the background spinors involve
only a sum of terms with two (or none) gamma matrices of \8M. It is then clear
that such projections will always commute with the F1 and the D0
ones, since they do not involve any gamma matrix of \8M.

To complete the proof, one must take into account further possible
problems that could be caused by the fact that the projections considered
so far are applied to background spinors which are not necessarily
constant. To see that this does not change the results, note
that~(\ref{constant}) implies that all the dependence of $\e$ on the \8M
coordinates $y^i$ must be of the form
\begin{equation}
\e=M(y)\e_0\,,
\end{equation}
with $\e_0$ a constant spinor, and $M(y^i)$ a matrix that involves only products of even number
of gamma matrices on \8M (it may well happen that $M(y)=\unity$).
Now, any projection on $\e$ can be translated to a projection on
$\e_0$ since
\[
P\e=\e\,, \espai \mbox{with} \espai P^2=\unity \,,\qquad  \trace P=0\,,
\espai \Longrightarrow
\]
\begin{equation}
\tilde{P}\e_0 =\e_0\,, \espai \mbox{with} \espai \tilde{P}\equiv M^{-1}(y) P
M(y)\,, \quad
\tilde{P}^2=\unity \,,\quad  \trace \tilde{P}=0\,.
\end{equation}
The only subtle point here is that, if some of the $\e_0$ have to
survive, the product of $ M^{-1}(y) P M(y)$ must be a constant
matrix\footnote{Note that it is not necessary that $P$ commutes with
$M(y)$.}. But this is always the case for all the projections
related to the presence of \8M, since we
know that such spaces preserve some Killing spinors.
Finally, it is also the case for the F1 and D0 projections, since
they commute with any even number of gamma matrices on \8M.

The conclusion is that, for an arbitrary curve in \8M to preserve
supersymmetry, it is necessary and sufficient to impose the F1 and D0
projections. In all cases, it will preserve $1/4$ of the background
supersymmetry. We will illustrate this with particular examples in
section~\ref{ss:examples}.

\avall

{\bf Non-Arbitrary Curve:} If we now give up the restriction that the
curve must be arbitrary, we can still show that the F1 and D0 projection
are necessary and sufficient, except for those cases in which the curve
lies entirely in the flat directions that \8M may have. Of course, the
former discussion shows that such projections are always sufficient, so
we will now study in which cases they are necessary as well.

In order to proceed, we need to prove an intermediate result.

\noindent {\it Lemma: If the velocity of the curve does not point in a
flat direction of \8M, then the background spinor always satisfies at
least one projection like
\begin{equation} \label{esquematic}
P\e=Q\e   \,, \qquad \mbox{such that}\qquad [P,\gamma_2]=0 \,,\qquad
\{Q,\gamma_2\}=0\,,
\end{equation}
with $P$ and $Q$ a non-vanishing sum of terms involving only an even
number of gamma matrices, and $Q$ invertible}.

To prove this, we move to a point of the curve that lies in a curved
direction of \8M, i.e. a point where not all components of
$R_{ij}{}^{\underline{kl}}$ are zero. We perform a rotation in the
tangent space such that the velocity of the curve points only in one of
the curved directions, {\it e.g.}
\begin{equation}
y'^{\underline{9}}\neq 0 \,, \espai \espai y'^{\underline{a}}=0 \,,
\espai \espai a=2,...,8 \,,\qquad R_{ij}{}^{a9}\neq 0\,,
\label{9direction}
\end{equation}
at least one choice of $i$, $j$ and $a$, and where we use the definitions
of~(\ref{gamma2}). With this choice, $\g_2$ becomes simply
$\gamma_2=y'^{\underline{9}} \G_{\underline{9}}$. Therefore, at least one
of the equations in~(\ref{integrability}) can be split in
\begin{equation}
\left( R_{ij}{}^{\underline{ab}}\G_{\underline{ab}} +
R_{ij}{}^{\underline{a9}}\G_{\underline{a9}} \right) \e=0 \,,
\end{equation}
with the definitions
\begin{equation} \label{define}
P= R_{ij}{}^{\underline{ab}}\G_{\underline{ab}} \,,  \espai\espai Q=
-R_{ij}{}^{\underline{a9}}\G_{\underline{a9}} \,.
\end{equation}
The assumption~(\ref{9direction}) implies that $Q$ is nonzero and
invertible, as the square of $Q$ is a negative definite multiple of the
unit matrix. This implies that also $P$ is non-zero since, otherwise,
$\e$ would have to be zero and this is against the fact that all the
listed \8M manifolds admit covariantly constant spinors. It is now
immediate to check that $\gamma_2$ commutes with $P$ while it
anticommutes with $Q$, which completes the proof. \QED

\avall

We can now apply this lemma and rewrite one of the conditions
in~(\ref{integrability}) as an equation of the kind~(\ref{esquematic}).
We then multiply the $\k$-symmetry condition~(\ref{ksym}) by $P-Q$.
Clearly only the first two terms survive, and we can write
\begin{equation}
0=\left[  \Gamma _{\underline{01}}-E \Gamma _*\right](P-Q)\gamma _2
\epsilon =- 2 \left[  \Gamma _{\underline{01}}-E \Gamma _*\right]\gamma
_2 Q\epsilon=-2\gamma _2 Q\left[  \Gamma _{\underline{01}}+E \Gamma
_*\right]\epsilon\,.
 \label{Pkappa}
\end{equation}
Since $(\gamma_2)^2=y'_iy'^i$ cannot be zero if the curve is not
degenerate, we just have to multiply with $Q^{-1}\gamma _2$ to find
again~(\ref{F1}). Plugging this back into~(\ref{ksym}) gives the
remaining D0 condition~(\ref{D0}).

\avall

Summarising, the usual supertube conditions are always necessary and
sufficient except for those cases where the curve is not required to be
arbitrary and lives entirely in flat space; then, they are just
sufficient. For example, one could choose $\CC$ to be a straight line in
one of the $\CR$ factors that some of the \8M have, and take a constant
$B$, which would correspond to a planar D2-brane preserving $1/2$ of the
background supersymmetry.

\section{Hamiltonian analysis}\label{ss:Hamilanal}

We showed that in order for the supertube
configurations~(\ref{embed}),~(\ref{fluxes}) to be supersymmetric we
needed $E^2=1$, but we found no restriction on the magnetic field
$B(\s^1,\s^2)$. We shall now check that some conditions must hold in
order to solve the equations of motion of the Maxwell fields. We will go
through the Hamiltonian analysis which will enable us to show that these
supertubes saturate a BPS bound which, in turn, implies the second-order
Lagrange equations of the submanifold determined by the constraints. We
will restrict to time-independent configurations, which we have checked
to be compatible with the full equations of motion. The Lagrangian is
then given by~(\ref{delta})
\begin{equation}
  {\cal L}=-\Delta =-\sqrt{B^2+R^2(1-E^2)}\,,
 \label{Lgeneral}
\end{equation}
where we have defined $R^2=y'^{\underline{i}}y'_{\underline{i}}$, and
$R>0$. To obtain the Hamiltonian we first need the displacement field,
\begin{equation}
  \Pi= \frac{\partial {\cal L}}{\partial
  E}\,=\frac{E R^2}{\sqrt{B^2+(1-E^2) R^2}}\,,
 \label{valuePi}
\end{equation}
which can be inverted to give
\begin{equation}
E=\frac{\Pi}{R} \sqrt{\frac{B^2+R^2}{R^2+\Pi ^2}}\,,\qquad \Delta=R
\sqrt{\frac{B^2+R^2}{ R^2+\Pi ^2}}\,.
 \label{E2}
\end{equation}

The Lagrange equations for $A_0$ and $A_2$ give two constraints
\begin{equation}
\partial_1 \Pi=0 \,, \espai\espai
  \partial _1\left(\frac{B}{ R}\sqrt{\frac{ R^2+\Pi ^2}{B^2+
  R^2}}\right)=0\,,
\label{gausslaw}
\end{equation}
the first one being the usual Gauss law. Together, they imply that
$\partial _1B=0$, i.e., the magnetic field can only depend on $\sigma^2$.
Finally, the equations for $A_1$ and $y^{\underline{i}}$ give,
respectively,
\begin{equation}
\partial _2\left(\frac{B}{R}\sqrt{\frac{ R^2+\Pi ^2}{B^2+
  R^2}}\right)=0
 \,, \qquad
 \partial _2\left[2y'^{\underline{i}} \frac{ R^4-\Pi ^2B^2}
 { R^2\sqrt{( R^2+\Pi ^2)( R^2+B^2)}} \right]=0 \,.
 \label{eom}
\end{equation}

The Hamiltonian density   is given by
\begin{equation}
  \mathcal{H}=E\Pi- \mathcal{L}=
\frac{1}{R}\sqrt{(R^2+\Pi ^2)(B^2+R^2)}\,.
 \label{valueH}
\end{equation}
In order to obtain a BPS bound~\cite{Gauntlett:1998ss}, we rewrite the
square of the Hamiltonian density as
\begin{equation}
  \mathcal{H}^2= \left(\Pi \pm B\right)^2+
  \left(\frac{\Pi B}{ R}\mp R\right)^2\,,
 \label{reH}
\end{equation}
from which we obtain the local inequality
\begin{equation} \label{inequality}
  \mathcal{H}\geq |\Pi\pm B|\,,
\end{equation}
which can be saturated only if
\begin{equation} \label{satura}
R^2=y'^{\underline{i}}y'_{\underline{i}}=\pm \Pi B \espai \Leftrightarrow
\espai E^2=1 \,.
\end{equation}
It can be checked that the configurations saturating this bound satisfy
the remaining equations of motion~(\ref{eom}).

Note that the Poynting vector generated by the electromagnetic field is
always tangent to the curve $\CC$ and its modulus is precisely $|\Pi B|$.
We can then use exactly the same arguments as in~\cite{Mateos:2001pi}.
Equation~(\ref{satura}) tells us that, once we set $E^2=1$, and
regardless of the value of $B(\s^2)$, the Poynting vector is
automatically adjusted to provide the required centripetal force that
compensates the gravitational attraction at every point of $\CC$. The
only difference with respect to the original supertubes in flat space is
that the curvature of the background is taken into account in
(\ref{satura}), through the explicit dependence of $R^2$ on the metric of
\8M.

Finally, the integrated version of the BPS bound~(\ref{inequality}) is
\begin{equation}
\tau \geq |q_0 \pm  q_s|\,, \espai \textrm{with} \espai \tau\equiv
\int_{\CC} {\rmd \sigma^2} \, \mathcal{H} \,,\qquad
q_0 \equiv \int_{\CC} {\rmd \sigma^2}\, B \,, \qquad
q_s \equiv \int_{\CC} {\rmd \sigma^2} \, \Pi \,.
\end{equation}
and the normalisation $0\leq \s^2 < 1$. Similarly, the integrated bound
is saturated when
\begin{equation} \label{length}
L(\CC)=\int_{\CC} {\rmd \sigma^2} \sqrt{  g_{22}}
 = \int_{\CC} {\rmd \sigma^2} \sqrt{y'^{\underline{i}}y'_{\underline{i}}}
 = \int_{\CC} {\rmd \sigma^2} \sqrt{|\Pi B|}
 =\sqrt{|q_s\,q_0|}\,,
\label{integratedbps}
\end{equation}
where $L(\CC)$ is precisely the proper length of the curve $\CC$,
and the last equality is only valid when both $\Pi$ and $B$ are constant,
as will be the case in our examples.

\section{Examples} \label{ss:examples}
After having discussed the general construction of supertubes in reduced
holonomy manifolds, we shall now present two examples in order to
illustrate some of their physical features.
\subsection{Supertubes in ALE spaces: 4 supercharges}

Let us choose \8M$=\CR^4 \times CY_2$, i.e. the full model being
$\mathbb{R}^{1,5}\times CY_2$. We take the $CY_2$ to be an ALE space
provided with a multi-Eguchi--Hanson metric~\cite{Eguchi:1978xp}
\begin{eqnarray}
&&\rmd s^2_{(4)}=V^{-1}(\yvec) \rmd\yvec \cdot \rmd\yvec +
V(\yvec)\left(\rmd\psi + \vec{A}\cdot \rmd\yvec\right)^2  \,, \nonumber\\
&&V^{-1}(\yvec)=\sum_{r=1}^{N} {Q\over |\yvec-\yvec_r|} \,,
\espai\espai\espai \vec{\nabla}\times \vec{A}=\vec{\nabla}V^{-1}(\yvec)
\,,
 \end{eqnarray}
with $\yvec \in \CR^3$. These metrics describe a $\U(1)$ fibration over
$\CR^3$, the circles being parametrized by $\psi\in [0,1]$. They present
$N$ removable bolt singularities at the points $\yvec_r$, where the
$\U(1)$ fibres contract to a point. Therefore, a segment connecting any
two such points, together with the fibre, form (topologically) an $S^2$.
For simplicity, we will just consider the two-monopoles case which,
without loss of generality, can be placed at $\yvec=\vec{0}$ and
$\yvec=(0,0,b)$. Therefore, the complete IIA background is
\begin{equation} \label{ALEIIA}
\rmd s^2_{IIA}=-(\rmd x^0)^2+(\rmd x^1)^2+...+(\rmd x^5)^2+ \rmd
s^2_{(4)} \,,
\end{equation}
with
\begin{equation}
V^{-1}(\yvec)={ Q \over |\yvec|} +{Q\over
|\yvec-(0,0,b)|} \,.
\end{equation}
Let us embed the D2 supertube in a way such that its longitudinal
direction lies in $\CR^5$ while its compact one wraps and $S^1$ inside
the $S^2$ that connects the two monopoles. More explicitly,
\begin{equation} \label{firstcase}
X^0=\sigma^0\,, \espai X^1=\sigma_1\,,\espai \psi=\sigma^2\,, \qquad
y^3=\textrm{const.} \,, \espai y^1=y^2=0 \,.
\end{equation}

\begin{figure}[here]
\begin{center}
\includegraphics[width=10.5cm,height=4.5cm]{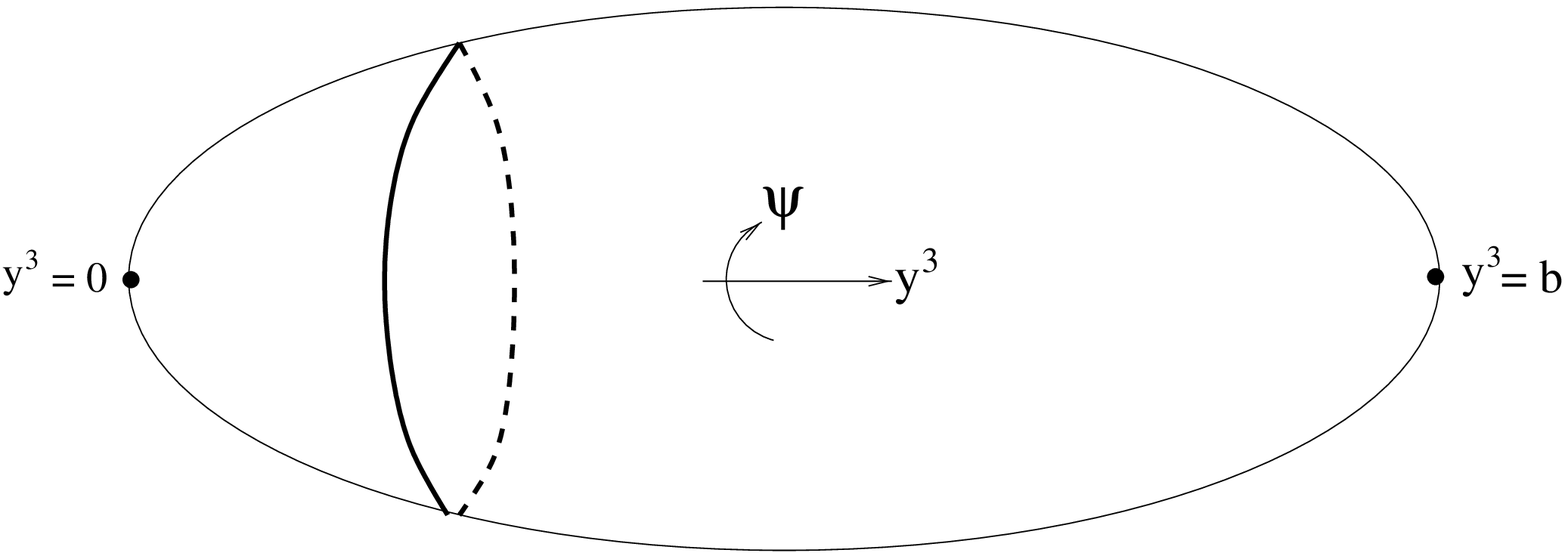}
\end{center}
\end{figure}
Since any $S^1$ is contractible inside an $S^2$, the curved part would
tend collapse to the nearest pole, located at $y^3=0$ or $y^3=b$. As in
flat space, we therefore need to turn on a worldvolume flux as
in~(\ref{fluxes}),
with $E$ and $B$ constant for the moment.

According to our general discussion, this configuration should
preserve $1/4$ of the 16 background supercharges already
preserved by the $ALE$ space. In this case, the $\k$-symmetry equation
is simply
\begin{equation}  \label{ALEkappa}
\left(\G_{\underline{01\psi}}+E \G_{\underline{\psi}}\Gamma
_*+B\G_{\underline{0}}\Gamma _* - \Delta \right)\e=0 \,,
\end{equation}
where $\e$ are the Killing spinors of the background~(\ref{ALEIIA}). They
can easily be computed and shown to be just constant spinors subject to
the projection
\begin{equation} \label{PALE}
\G_{\underline{y^1y^2y^3\psi}}\e=-\e \,.
\end{equation}
Then, the $\kappa $-symmetry equation can be solved by
requiring~(\ref{F1}) and~(\ref{D0}),
which involve the usual D0/F1 projections of the supertube. Since they
commute with~(\ref{PALE}), the configuration preserves a total of $1/8$
of the 32 supercharges.

It is interesting to see what are the consequences of having $E^2=1$ for
this case. Note that, from our general Hamiltonian analysis, we saw that,
for fixed D0 and F1 charges, the energy is minimised for $E^2=1$. When
applied to the present configuration,~(\ref{length}) reads
\begin{equation} \label{selects}
V(y^3)=|q_0 q_s| \,.
\end{equation}
which determines $y^3$, and therefore selects the position of
the $S^1$ inside the $S^2$ that is compatible with supersymmetry.
Since $V(y^3)$ is invariant under $y^3 \leftrightarrow (b-y^3)$,
the solutions always come in mirror pairs with respect to the
equator of the $S^2$. The explicit solutions are indeed
\begin{equation}
y^3_{\pm}={b\over 2} \left(1 \pm \sqrt{1-{4Q\over b}|q_0q_s|}\right)
\;.\end{equation} Note that a solution exists as long as the product of
the charges is bounded from above to
\begin{equation} \label{solu}
|q_0 q_s| \leq {b\over 4Q} \;.
\end{equation}
The point is that this will always happen due to the fact that,
contrary to the flat space case, the $S^1$ cannot grow arbitrarily
within the $S^2$. As a consequence, the angular momentum acquires its
maximum value when the $S^1$ is precisely in the equator. To see
it more explicitly, setting $E^2=1$ and computing
$q_0$ and $q_s$ for our configuration gives
\begin{equation}
|q_0 q_s| = V(y^3) \leq V(y^3 \rightarrow {b\over 2}) ={b\over 4Q} \;,
\end{equation}
which guarantees that~(\ref{solu}) is always satisfied.

Finally, note that we could have perfectly chosen, for instance, a more
sophisticated embedding in which $y^3$ was not constant. This would be
the analogue of taking a non-constant radius in the original flat space
supertube. Again, by the general analysis of the previous sections, this
would require the Poynting vector to vary in order to locally compensate
for the gravitational attraction everywhere, and no further supersymmetry
would be broken.

\subsection{Supertubes in $CY_4$ spaces: 1 supercharge}

The purpose of the next example is to show how one can reach a
configuration with one single surviving supercharge in a concrete
example. One could take any of the $1/8$-preserving backgrounds of the
\8M Table. Many metrics for these spaces have been recently found in the
context of supergravity duals of non-maximally supersymmetric field
theories. Let us take the $CY_4$ that was found
in~\cite{Gomis:2001vg,Cvetic:2000db} since the Killing spinors have been
already calculated explicitly~\cite{Brugues:2002ff}. This space is a
$C^2$ bundle over $S^2\times S^2$, and the metric is
\begin{eqnarray}
  \rmd s^2_{(CY_4)}&=&A(r)\left[ \rmd \theta _1^2+\sin^2\theta _1\rmd
\phi _1^2+
\rmd \theta _2^2+\sin^2\theta _2\rmd
\phi _2^2
\right]
  +U^{-1}\rmd r^2+ {r^2 \over 4}\left( \rmd\theta^2+ \sin^2\theta
\rmd\phi^2\right)+  \nonumber\\ && +{1\over 4}U r^2\left( \rmd\psi+\cos\theta \rmd\phi
+\cos\theta _1\rmd \phi _1+\cos\theta _2\rmd \phi _2\right) ^2\,,
 \label{metric11}
\end{eqnarray}
where
\begin{equation}
  A(r)={3\over 2}(r^2+l^2)\,, \qquad
  U(r)={3 r^4 + 8 l^2 r^2 + 6 l^4 \over {6(r^2+l^2)^2}} \,,\qquad
  C(r)=\frac14 U\,r^2\,.
 \label{defA}
\end{equation}
By writing the complete IIA background metric as
\begin{equation}
\rmd s^2_{IIA}=-(\rmd x^0)^2 +(\rmd x^1)^2 + \rmd s^2_{(CY_4)}\,,
\end{equation}
and using the obvious vielbeins, with the order
\begin{equation}
   \begin{array}{cccccccc}
    2& 3 & 4 & 5 & 6 & 7 & 8 & 9  \\
     \theta _1 & \theta _2 & \phi _2 & \phi _1 & r & \theta  & \phi  &
     \psi
  \end{array}
 \label{variables}
\end{equation}
the corresponding Killing spinors are
\begin{equation}
  \epsilon =\rme^{-\ft12\psi \Gamma _{\underline{78}}}\epsilon _0\,,
\end{equation}
with $\e_0$ a constant spinor subject to
\begin{equation}
 \Gamma _{\underline{25}}\epsilon_0 =\Gamma _{\underline{34}}\epsilon_0\,,\qquad
 \Gamma _{\underline{25}}\epsilon_0 =\Gamma_{\underline{78}}\epsilon_0 \,,\qquad
 \Gamma _{\underline{67}}\epsilon_0 =\Gamma _{\underline{98}}\epsilon_0 \,.
 \label{projCY4}
\end{equation}
To analyse $\k$-symmetry, let us take
the compact part of the supertube to lie along, say, the $\phi_1$
direction, while setting to constant the rest of the $CY_4$ coordinates.
As in the previous example, this would have the interpretation of
an $S^1$ embedding in one of the two $S^2$ in the base of the $CY_4$. Imposing
$\k$-symmetry:
\begin{equation}  \label{CY4kappa}
\left(\G_{\underline{015}}+E \G_{\underline{5}}\Gamma
_*+B\G_{\underline{0}}\Gamma _* - \Delta \right)\e=0 \,.
\end{equation}
Now, the first projection of~(\ref{projCY4}) happens to anticommute with
the $\gamma_2$ defined in~(\ref{gamma2})
\begin{equation}
\gamma _2=y'^ie_i{}^{\underline{i}}\Gamma _{\underline{i}}\, =\,
A^{1\over 2}(r)\, \sin{\theta_1} \, \Gamma_{\underline{5}} \,.
\end{equation}
In other words, this just illustrates a particular case
of~(\ref{esquematic}) for which the direction $\underline{5}$ plays the
role of $\underline{9}$, and for which $P=\Gamma _{\underline{34}}$ and
$Q=\Gamma _{\underline{25}}$. We can now follow the steps in
section~\ref{ss:proof} and multiply~(\ref{CY4kappa}) by $P-Q$. This
yields again the usual supertube conditions~(\ref{F1}) and~(\ref{D0}).

Since all the gamma matrices appearing in~(\ref{projCY4}),~(\ref{F1})
and~(\ref{D0}) commute, square to one and are traceless, the
configuration preserves only one of the 32 supercharges of the theory. Of
course, this is not in contradiction with the fact that the minimal
spinors in 2+1 dimensions have 2 components, since the field theory on
the worldvolume of the D2 is not Lorentz invariant because of the
non-vanishing electromagnetic field.

\section{Supergravity analysis}\label{ss:SGanal}

In this section we construct the supergravity family of solutions that
correspond to all the configurations studied before. We start our work
with a generalisation of the ansatz used
in~\cite{Emparan:2001ux,Mateos:2001pi} to find the original solutions.
Our analysis is performed in eleven dimensional supergravity, mainly
because its field content is much simpler than in IIA supergravity.
Once the eleven-dimensional solution is found,
we reduce back to ten dimensions, obtaining our generalised supertube
configurations.

The first step in finding the solutions is to look for supergravity
configurations with the isometries and supersymmetries suggested by the
worldvolume analysis of the previous sections. Then, we will turn to the
supergravity field equations to find the constraints that the functions
of our ansatz have to satisfy in order that our
configurations correspond to minima of the eleventh dimensional
 action.
Finally, we choose the correct behaviour for these functions so that they
correctly describe the supertubes once the reduction to ten dimensions is
carried on.

\subsection{Supersymmetry analysis}

Our starting point is the supertube ansatz
of~\cite{Emparan:2001ux,Mateos:2001pi}
 \begin{eqnarray}
  \rmd s^2_{\it 10} &=& - U^{-1}
V^{-1/2} \, ( \rmd t - A)^2 + U^{-1} V^{1/2} \, \rmd x^2 + V^{1/2} \,
\delta_{ij}\rmd y^i\rmd y^j \,, \nn B_{\it 2} &=& - U^{-1} \, (\rmd t -
A) \wedge \rmd x + \rmd t\wedge \rmd x\,, \nn C_{\it 1} &=& - V^{-1} \,
(\rmd t - A) + \rmd t \,, \nn C_{\it 3} &=& - U^{-1} \rmd t\wedge \rmd x
\wedge A \,, \nn e^\phi &=& U^{-1/2} V^{3/4} \,, \label{ds10}
 \end{eqnarray}
where the Euclidean space ($\mathbb{E}_8$) coordinates are labelled by
$y^i$, with $i,j,\cdots = (2,\ldots,9)$, $V=1+K$, $A=A_i\,\rmd y^i$ and
$B_{\it 2}$ and $C_{\it p}$ are respectively, the Neveu-Schwarz and
Ramond-Ramond potentials. $V,U,A_i$ depend only on the $\mathbb{E}_8$
coordinates.

To up-lift this ansatz, we use the normal Kaluza-Klein form of the eleven
dimensional metric and three-form,
\begin{eqnarray}
\rmd s^2_{\it 11} &=& e^{-2\phi/3}\rmd s^2_{\it 10}+e^{4\phi/3}(\rmd
z+C_{\it 1})^2 \,,\nonumber\\
N_{\it 3 } &=& C_{\it 3}+B_{\it 2} \wedge \rmd z \,,\label{uplift}
\end{eqnarray}
where $N_{\it 3 }$ is the eleventh dimensional three-form.
The convention for curved indices is $M=(\mu;i)=(t,z,x\,;\,2,3,...9)$ and
for flat ones
$A=(\alpha;a)=(\underline{t},\underline{z},\underline{x}\,;\,\underline{2},\underline{3}...,\underline{9})$.
The explicit form of the eleven-dimensional metric is given by,
\begin{eqnarray}
&&\rmd s^2_{\it 11} = U^{-2/3}\left[ -\rmd t^2 + \rmd z^2 + K(\rmd t+\rmd z)^2
+ 2(\rmd t+\rmd z)A +\rmd x^2\right]+U^{1/3}\rmd s^2_{\it 8}\,, \nonumber \\
&&F_{\it 4} = \rmd t \wedge \rmd(U^{-1})\wedge \rmd x\wedge \rmd z -
(\rmd t+\rmd z)\wedge \rmd x \wedge \rmd(U^{-1}A)\;, \label{ansatz}
\end{eqnarray}
where $F_{\it 4} = \rmd N_{\it 3}$. This background is a solution of the
equations of motion in eleven dimensions derived from the action
\begin{equation}
S_{11d}=\int \,\left[ R *1 \, - \, {1\over 2} F_{\it 4} \wedge *
 F_{\it 4} \, + \, {1 \over 3}
F_{\it 4} \wedge F_{\it 4} \wedge N_{\it 3} \right] \,,
\end{equation}
when the two functions $K$ and $U$, as well as the one-form $A_{\it 1}$,
are harmonic in $\mathbb{E}_8$, i.e.,
\begin{equation}
(\rmd *_8\rmd)U=0 \,, \zespai (\rmd *_8\rmd)K=0 \,, \zespai (\rmd
*_8\rmd)A_{\it 1}=0 \,, \label{harmonic}
\end{equation}
where $*_8$ is the Hodge dual with respect to the Euclidean flat metric
on $\mathbb{E}^8$. It describes a background with an M2 brane along the
directions $\{t,z,x\}$, together with a wave traveling along $z$, and
angular momentum along $\euc^8$  provided by $A_{\it 1}$.

Next, we generalise the ansatz above by replacing $\euc^8$ by one of the
eight dimensional \8M manifolds of the table, and by allowing $K$, $U$
and $A_{\it 1}$ to have an arbitrary dependence on the \8M coordinates
$y^i$. We therefore replace the previously flat metric on $\euc^8$ by a
reduced holonomy metric on \8M, with vielbeins $\te^a$. Hence, in
(\ref{ansatz}), we replace
\begin{equation}
U^{1/3} \delta_{ij} \rmd y^i \rmd y^j \espai \longrightarrow \espai
U^{1/3} \delta_{ab}\te^a \te^b  \,.\label{E8byM8}
\end{equation}

We use a null base of the cotangent space, defined by
\begin{eqnarray}
\label{secondbase}
&& e^+=-U^{-2/3}(\rmd t+\rmd z) \,, \zespai e^-=\undos (\rmd t-\rmd z) -
{K\over 2} (\rmd t+\rmd z) - A \,,\nonumber\\
&& e^x=U^{-1/3} \rmd x  \,, \zespai e^a=U^{1/6} \te^a \,.
\end{eqnarray}
This brings the metric and $F_{\it 4}$ into the form
\begin{equation} \label{flat}
\rmd s^2_{\it 11}=2 e^+ e^- + e^x e^x + \delta_{ab}e^a e^b  \,, \zespai
F_{\it 4}=-U^{-1} \, \rmd U\we e^x \we e^+ \we e^- \, -\, \rmd A \we e^x
\we e^+ \,.
\end{equation}
As customary, the torsion-less condition can be used to determine the
spin connection 1-form $\omega _{AB}$. In our null base, the only
non-zero
components are 
\begin{eqnarray}
&&\omega_{+-}=-{U_a \over 3U} e^a \,, \zespai \omega_{+a}=\frac{1}{2}
U^{1/2}\tilde K_a e^+-{U_a \over 3U}e^- -\frac12a_{ab} e^b\,, \zespai
\omega_{-a}=-{U_a
\over 3U} e^+ \,, \nonumber\\
&&\omega_{xa}=-{U_a \over 3U} e^x \,,\qquad  \omega_{ab}={U_b \over
6U}e^a -{U^a \over 6U}e^b +\tilde \omega _{ab} +\frac12a_{ab} e^+ \,,
\label{spin2}
\end{eqnarray}
were we have defined various tensor quantities through the relations
\begin{equation}
\rmd U = U_a e^a \,, \zespai  \rmd K=\tilde K_a \tilde e^a \,, \zespai
\rmd A=\ft12a_{ab} e^a \we e^b\,,
\end{equation}
and $\tilde{\omega}^{bc}$ are the spin connection one-forms corresponding
to $\tilde{e}^a$, i.e. $\rmd \tilde e^a+\tilde \omega ^a{}_b\tilde
e^b=0$.

We now want to see under which circumstances our backgrounds preserve
some supersymmetry. Since we are in a bosonic background i.e. all the
fermions are set to zero, we just need to ensure that the variation of
the gravitino vanishes when evaluated on our configurations. In other
words, supersymmetry is preserved if there exist nonzero background
spinors $\e$ such that\footnote{For the components of $p$-forms we use
the notations of~\cite{Candelas:1987is}.}
\begin{equation} \label{gravitino}
\left(\p_A+{1\over 4}\omega_A{}^{BC}\G_{BC} -{1\over
288}\G_A{}^{BCDE}F_{BCDE}+{1\over 36} F_{ABCD}\G^{BCD}\right)\e=0 \,.
\end{equation}
We will try an ansatz such that the spinor depends only on the
coordinates on \8M. It is straightforward to write down the eleven
equations~(\ref{gravitino}) for each value of $A=\{+,-,x,a\}$. The
equation for $A=x$ is
\begin{equation}
 {U_a \over 6U} \Gamma _a\left(\Gamma_{x}-\Gamma_{+-}\right) \epsilon
 -{a_{ab}\over 12}\Gamma_{ab}\Gamma _-\epsilon =0\,.
 \label{SUSYx}
\end{equation}
Assuming that $a_{ab}$ and $\alpha _a$ are arbitrary and independent we
find
\begin{equation} \label{proj1}
\G_- \, \e=0 \,, \zespai \textrm{and} \zespai \G_{x}\e=-\e \,.
\end{equation}
Using these projections, it is a straightforward algebraic work to see
that the equation for $A=+$ and $A=-$ are automatically satisfied.
Finally, the equations for $A=a$ simplify to
\begin{equation} \label{proj2}
\nabla_{i}\e \, \equiv \, \left(\p_{i}+{1\over 4}
\tilde{\omega}_i{}^{bc}\G_{bc}\right)\e =0 \,.
\end{equation}

By the same arguments as in the previous sections, the
projections~(\ref{proj1}) preserve 1/4 of the 32 real supercharges. On
the other hand,~(\ref{proj2}) is just the statement that \8M must admit
covariantly constant spinors. Depending on the choice of \8M, the whole
11d background will preserve the expected total number of supersymmetries
that we indicated in the table written in the introduction.

To reduce back to IIA supergravity, we first go to another flat basis
\begin{equation}
  e^+= -U^{-1/3}V^{-1/2}\left( e^0+e^z\right)\,,\qquad
e^-=\ft12 U^{1/3}V^{1/2}\left( e^0-e^z\right)\,,
 \label{e+-tz}
\end{equation}
which implies that
\begin{equation}
  \Gamma _-=U^{-1/3}V^{-1/2}\left( \Gamma _0-\Gamma _z\right).
 \label{Gamma-zt}
\end{equation}
We reduce along $z$, i.e.\ replace $\Gamma _z$ by $\Gamma _*$. The
projections~(\ref{proj1}) become the usual D0/F1 projections, with the
fundamental strings along the $x$-axis.
\begin{equation}
\G_0\Gamma _*\e=-\e  \,, \zespai \textrm{and} \zespai \epsilon =-\Gamma
_x\epsilon =\G_{x0}\Gamma _*\e \,.
\end{equation}

\subsection{Equations of motion}

Now that we have proved that the correct supersymmetry is preserved
(matching the worldvolume analysis), we proceed to determine the
equations that $U$, $K$ and $A_{\it 1}$ have to satisfy in order that our
configurations solve the field equations of eleven-dimensional
supergravity. Instead of checking each of the equations of motion, we use
the analysis of~\cite{Gauntlett:2002fz} that is based on the
integrability condition derived from the supersymmetry variation of the
gravitino~(\ref{gravitino}). The result of this analysis is that when at
least one supersymmetry is preserved, and the Killing vector
$\mathcal{K}_\mu \equiv \bar \epsilon \Gamma_\mu \epsilon$ is null, all
of the second order equations of motion are automatically satisfied,
except for
\begin{enumerate}
  \item The equation of motion for $F_{\it 4}$,
  \item The Einstein equation $E_{++}=T_{++}$,
\end{enumerate}
where $E_{++}$ and $T_{++}$ are the Einstein and stress-energy tensors
along the components $++$ in a base where $\mathcal{K}_\mu=\delta _\mu
^+\mathcal{K}_+$. Let us explain why the above statement is correct. The
integrability conditions give no information about the field equation for
the matter content, therefore the equation of motion for $F_{\it 4}$ has
to be verified by hand. Also, in most cases all of the Einstein equations
are automatically implied by the existence of a non-trivial solution
of~(\ref{gravitino}).

With~(\ref{proj1}) and in the base where the metric takes the
form~(\ref{flat}), and thus $\Gamma _+\Gamma _-+\Gamma _-\Gamma _+=2$, we
have
\begin{equation}
  \mathcal{K}_\mu =\bar \epsilon \Gamma_\mu \epsilon=\ft12\bar \epsilon \Gamma_\mu \Gamma _-\Gamma
  _+\epsilon\,.
 \label{Kmu}
\end{equation}
This vanishes for all $\mu $ except $\mu =+$, implying that our
configuration falls into the classification of those backgrounds that
admit a null Killing spinor and as a consequence the associated Einstein
equations escape the analysis. 
We thus have to check the two items mentioned above.

Let us start with the equation for $F_{\it 4}$, which is
\begin{equation} \label{fourform}
\rmd *F_{\it 4} + F_{\it 4} \we F_{\it 4} = 0 \,.
\end{equation}
Using the fact that the Hodge dual of a p-form
with respect to $e^a$ is related to the one with respect to
$\te^a$ by
\begin{equation}
*_8 C_p = U^{(4-p)/3} \t*_8 C_{p} \,,
\end{equation}
where
\begin{equation}
C_p=\frac{1}{p!}C_{a_1\ldots a_p}\tilde e^{a_1}\wedge \ldots \wedge
\tilde e^{a_p}\ \rightarrow \tilde *_8 C_{p}=
\frac{1}{p!(8-p)!}C_{a_1\ldots a_p} \varepsilon ^{a_1\ldots a_8}\tilde
e^{a_{p+1}}\wedge \ldots \wedge \tilde e^{a_8}\,,
\end{equation}
it is easy to see that~(\ref{fourform}) becomes
\begin{equation}
0= (\rmd\t*_8\rmd) U + (\rmd t+\rmd z)\we (\rmd\t*_8\rmd)A \,.
\end{equation}
This implies that $U$ and $A_{\it 1}$ must be harmonic with respect
to the metric of \8M, i.e.,
\begin{equation}
(\rmd\t*_8\rmd) U = 0 \,, \zespai (\rmd\t*_8\rmd)A_{\it 1}=0 \,.
\label{harmonicUA}
\end{equation}
Finally, using~(\ref{flat}) and~(\ref{spin2}), one can explicitly compute
the $\{++\}$ components of the Einstein and stress-energy tensors, and
obtain
\begin{eqnarray}
E_{++}&=&R_{++}=-\ft12U^{1/3}(\t*_8\rmd\t*_8\rmd)K + \ft12 *_8\left(\rmd
A\we *_8 \rmd A\right) \,,\nonumber\\
T_{++}&=&\ft1{12}F_{+ABC}F_+{}^{ABC} = \ft12 *_8\left(\rmd A\we *_8 \rmd
A\right) \,,
\end{eqnarray}
Therefore, the last non-trivial equation of motion tells us
that also $K$ must be harmonic on \8M,
\begin{equation}
(\rmd\t*_8\rmd)K=0 \,. \label{harmonicK}
\end{equation}

\subsection{Constructing the supertube}

In order to construct the supergravity solutions that properly describe
supertubes in reduced holonomy manifolds, we reduce our
eleven-dimensional background to a ten-dimensional background of type IIA
supergravity, using~(\ref{uplift}) again. We obtain~(\ref{ds10}) with the
replacement~(\ref{E8byM8}), and the constraints~(\ref{harmonicUA})
and~(\ref{harmonicK}).
At this point we have to choose $U$, $K$ and $A_{\it 1}$ so that they
describe a D2-brane with worldvolume $\CR^{1,1} \times \CC$, with $\CC$
an arbitrary curve in \8M. As it was done
in~\cite{Emparan:2001ux,Mateos:2001pi}, one should couple IIA
supergravity to a source with support along $\CR^{1,1} \times \CC$, and
solve the \8M Laplace equations~(\ref{harmonicUA}) and~(\ref{harmonicK})
with such a source term in the right hand sides. If this has to
correspond to the picture of D0/F1 bound states expanded into a D2 by
rotation, the boundary conditions of the Laplace equations must be such
that the solution carries the right conserved charges. In the appropriate
units,
\begin{equation}
q_0=\int_{\p \CM_8} \t*_8 \rmd C_{\it 1} \,, \zespai q_s=\int_{\p \CM_8}
\t*_8 \rmd B_{\it 2} \,, \zespai A_{\it 1}
 \stackrel{\p \CM_8}{\longrightarrow}L_{ij}y^j \rmd y^i \,.
\end{equation}
Here, as in~\cite{Emparan:2001ux,Mateos:2001pi}, $L_{ij}$ would have to
match with the angular momentum carried by the electromagnetic field that
we considered in the worldvolume approach.

The Laplace problem in a general manifold can be very complicated and, in
most cases, it cannot be solved in terms of ordinary functions. We will
not intend to do so, but rather we will just claim that, once $U$, $K$
and $A_{\it 1}$ have been determined, they can be plugged back
into~(\ref{ds10}), with~(\ref{E8byM8}), and the background will describe
the configurations that we have been discussing in this paper. It will
have the expected isometries, supersymmetries and conserved charges.

\section{Conclusions}\label{ss:concl}
We have shown that the expansion of the D0/F1 system into a D2 can happen
supersymmetrically in all the backgrounds of the form $\CR^{1,1} \times
\CM_8$, with \8M the manifolds of the table. We have shown this in the
worldvolume as well as in the supergravity setting. By a Hamiltonian
analysis, we connected the result to a BPS bound on charges that are also
well defined in the curved background. We remark that our research is
different from~\cite{Grandi:2002gt}, where it was shown that {\it the
supertube itself}, after some T-dualities, can be described by a special
Lorentzian-holonomy manifold in eleven dimensions.

\medskip
\section*{Acknowledgments.}

\noindent We are grateful to Roberto Empar{\'a}n, David Mateos, Guillermo A.
Silva, Joan Sim{\'o}n and Paul Townsend  for useful discussions. Work
supported in part by the European Community's Human Potential Programme
under contract HPRN-CT-2000-00131 Quantum Spacetime, in which P. Silva is
associated with Torino Universit{\`a}. This work is also supported by MCYT
FPA, 2001-3598 and CIRIT GC 2001SGR-00065. T. Mateos is supported by the
grant FI from the Generalitat de Catalunya. The work of A.V.P. is
supported in part by the Federal Office for Scientific, Technical and
Cultural Affairs through the Inter-university Attraction Pole P5/27.
\newpage
%
%

\providecommand{\href}[2]{#2}\begingroup\raggedright\endgroup

\end{document}